\documentstyle{article}

\newcommand{\be}{\begin{equation}}
\newcommand{\ee}{\end{equation}}
\newcommand{\bea}{\begin{eqnarray}}
\newcommand{\eea}{\end{eqnarray}}
\newcommand{\bean}{\begin{eqnarray*}}
\newcommand{\eean}{\end{eqnarray*}}

\newcommand{\F}{\mbox{\boldmath$ F$}}

\renewcommand{\u}{\mbox{\boldmath$ u$}}

\newcommand{\cK}{{\cal K}}
\newcommand{\cE}{{\cal E}}
\newcommand{\cF}{{\cal F}}
\newcommand{\cG}{{\cal G}}
\newcommand{\cU}{{\cal U}}
\newcommand{\cO}{{\cal O}}

\newcommand{\pard}[2]{\frac{\partial {#1}}{\partial {#2}}}

\newcommand{\cL}{{\cal L}}
\newcommand{\bs}{\mbox{\boldmath$ s$}}
\newcommand{\cH}{{\cal H}}
\newcommand{\cV}{{\cal V}}
\newcommand{\avu}{{\bar u}}
\newcommand{\avk}{{\bar k}}
\newcommand{\ave}{{\bar \epsilon}}

\newcommand{\sfrac}[2]{{\mbox{$\frac{#1}{#2}$}}}

\begin{document}

\title{EQUATIONS FOR TURBULENT FLOOD WAVES}
\author{Z.~MEI\thanks{Department of Mathematics, University of Marburg,
35032 Marburg/Lahn, Germany}
\and A.J.~ROBERTS\thanks{Department of Mathematics \& Computing,
University of Southern Queensland,
Toowoomba, Queensland 4350, Australia.}}
\date{4 Oct 1994}
\maketitle

\begin{abstract}
\footnotesize Long waves in rivers, estuaries and floods are described by the
St~Venant and Boussinesq equations in classical fluid dynamics. Based
on the widely used $k$-$\epsilon$ model for turbulence, we use the
techniques of centre manifold theory to show how these classical models can
be derived from a unified approach to low-dimensional modelling.
The model may be refined to an arbitrary degree of accuracy and predicts
interesting interactions between the mean flow and the mean turbulence.
\end{abstract}

\section{Introduction}

Modelling turbulent flow is a major challenge in fluid dynamics. While
large eddies, which can be as large as the flow domain, extract energy from
the mean flow and feed it into large scale turbulent motion, the eddies
also act as a vortex stretching mechanism and transfer the energy to the
smallest scales where viscous dissipation takes place. However, the inflow
of energy into turbulent motion is a characteristic of only the large scale
motion. In other words, the turbulent but small scale motion is often
dominated and determined by the large scale motion and can be treated as a
perturbation of the mean flow. The coupling of energy transportation and
energy dissipation with the mean flow is described by widely used
second-order closure models. In particular, the most popular choice is the
two equation $k$-$\epsilon$ model, see for example Launder
{\em et al}\cite{LRR1}, Hanjali\'c \& Launder\cite{HanLau},
Rodi\cite{Rodi}, and
Speziale\cite{Spez1} for reviews.

The recent development of centre manifold theory and related techniques
presages a much deeper understanding of the process of modelling and
foresees the systematic reduction of many nonlinear problems. In the
context of turbulent flow, it shows that even mean motions may be
mostly determined by a few critical modes and thus can be described by a lower
dimensional system of amplitude equations for these modes.

In this paper we use the $k$-$\epsilon$ model for the turbulence underlying
the free surface of fluid in a channel or river, \S2. We use centre
manifold techniques (\S3) to derive a model for the evolution of vertically
averaged quantities. The model resolves large-scale dynamical structures in
the horizontal, large compared to the depth. It should be useful in a wide
variety of environmental flows and may be compared to experiments in
channels with variable depth such as those by Shiono \& Knight\cite{Shiono91}.
Importantly, based on this theoretical background, one can
refine the various models to an arbitrary degree of accuracy for a given
problem. We show how classical models for long waves, for example the
St~Venant and Boussinesq equations (\S4), are derived as special cases of
the centre manifold reduction.

\section{The $k$-$\epsilon$ model of turbulent flow}

Consider the two-dimensional inviscid $k$-$\epsilon$ model of turbulent
flow over rough ground.  Distance parallel to the ground's slope is
measured by $x$, while we denote $y$ as
the direction normal to the slope. Molecular dissipation is
neglected because we see little role for it in flood waves of a depth
$h=\cO(\mbox{metres})$ over ground with roughness which may be many times the
length-scale of molecular dissipation. Turbulent eddies are proposed
to be the dominant mechanism for dispersion.  We denote the ensemble mean
velocity
components and pressure by $u$, $v$ and $p$ respectively, that is, for
simplicity, we omit any distinguishing overbars. Then the incompressible
$k$-$\epsilon$ model (in terms of ensemble means) may be expressed as
\be \label{EAKeFW}
\left[\begin{array}{c}0\\\pard{\u}{t}\end{array}\right]=\left[
\begin{array}{c}\pard{u}{x}+\pard{v}{y}\\ \F(p, \u)
\end{array}\right],
\ee
where the vector $\u:=(u,v, \eta, k,\epsilon)$ \footnote{We adopt the
notation that a vector in parentheses, such as $(u,v, \eta, k,\epsilon)$,
is a short-hand for the corresponding column vector.} is formed from the
velocities $u$, $v$ in the horizontal and vertical directions, the height of
the free surface $y=\eta(x,t)$, the turbulent energy density $k$, and its
dissipation rate $\epsilon$.  The nonlinear model governing the evolution
of the unknowns in $\u$ is
\begin{equation}
\F(p, \u)=
\left[ \begin{array}{c}
-u\pard{u}{x}-v\pard{u}{y}-\pard{p}{x}+g\sin\theta-\frac23\pard{k}{x}
+2\pard{~}{x}\left(\nu\pard{u}{x}\right)+\pard{~}{y}\left[\nu\left(
\pard{u}{y}+\pard{v}{x}\right)\right]\cr
-u\pard{v}{x}-v\pard{v}{y}-\pard{p}{y}-g\cos\theta-\frac23\pard{k}{y}
+2\pard{~}{y}\left(\nu\pard{v}{y}\right)+\pard{~}{x}\left[\nu\left(
\pard{u}{y}+\pard{v}{x}\right)\right]\cr
-u(x,\eta,t)\pard{\eta}{x}+v(x,\eta,t) \cr
-u\pard{k}{x}-v\pard{k}{y}+\left[
\pard{~}{x}\left({\nu\over\sigma_k}\pard{k}{x}\right)+
\pard{~}{y}\left({\nu\over\sigma_k}\pard{k}{y}\right)\right]+P_h
-\epsilon\cr -u\pard{\epsilon}{x}-v\pard{\epsilon}{y}+ \left[
\pard{~}{x}\left({\nu\over\sigma_\epsilon}\pard{\epsilon}{x}\right)+
\pard{~}{y}\left({\nu\over\sigma_\epsilon}\pard{\epsilon}{y}\right)\right]+
C_{\epsilon1}{\epsilon\over k}P_h
-C_{\epsilon2}{\epsilon^2\over k} \end{array} \right]\,.
\label{frhs}
\end{equation}
Here the eddy viscosity
 \begin{equation}
         \nu=C_{\mu}{k^2\over \epsilon}
         \label{nudef}
 \end{equation}
is a result of the turbulent mixing and varies in space and time, and
the following approximate values of the constants
$$
   C_\mu=0.09, ~~\sigma_k=1,~~\sigma_\epsilon=1.3,~~C_{\epsilon1}=1.44,
   ~~C_{\epsilon2}=1.92\,,
$$
are often used.  The term
$$
P_h=\nu\left[2\left(\pard{u}{x}\right)^2+2\left(\pard{v}{y}\right)^2
+\left(\pard{u}{y}+\pard{v}{x}\right)^2\right]
$$
describes the generation of turbulence through instabilities associated
with mean-velocity gradients. The slope of the bottom $\theta$ is assumed
to be small and to have negligible variation.

Boundary conditions on the bed and the fluid surface are
important in the construction of the low-dimensional model. The following
arguments lead to the given boundary conditions.
\begin{itemize}
\item A standard condition is that, in view of the extremely low density
of air, the pressure of the air on the fluid surface is constant which we
take to be zero.  Thus the normal stress of the fluid across the
free surface should vanish on the free surface:
\[
p+\frac{2}{3}k-\frac{2\nu}{1+\eta_x^2}\left[ \pard vy+\pard ux
-\eta_x\left( \pard vx+\pard uy\right)\right]=0\quad\mbox{on $y=\eta$.}
\]
\item In this work we assume that the horizontal extent of the flood
waves is small enough so that  wind stress is negligible.  This is in
contrast to large-scale geophysical simulations, such as that by
Arnold \& Noye\cite{AnNo82}, where the wind stress is very important.
Thus the fluid surface is free of tangential stress:
\[
\left(1-\eta_x^2\right)\left(\pard uy+\pard vx\right)
+2\eta_x\left(\pard vy-\pard ux\right)=0\quad\mbox{on $y=\eta$.}
\]
\item Symmetry conditions for $k$ and $\epsilon$ on the free surface
(e.g.~see Arnold \& Noye\cite{AnNo84}) lead to
\[
    \pard{k}{n}(x,\eta,t)=0,
    \quad\mbox{and}\quad\pard{\epsilon}{n}(x,\eta,t)=0.
\]
\item At the ground, $y=0$, there can be no flow across the flat bottom:
 \[
v(x,0,t)=0.
\]
\item Other boundary conditions on the ground are more arguable (compare
our treatment with that of Arnold \& Noye\cite{AnNo84,AnNo86}). We are only
interested in the flow outside of any molecular boundary layer that may
exist on the stream bed---the ground will be too rough for molecular
mechanisms to be important directly on the mean flow. On the bottom we
suppose that the ensemble mean velocity, $u$, interacts with the rough
ground to generate turbulent eddies with the same characteristic velocity:
$\sqrt k\propto u$. Thus we would impose the boundary condition
\begin{equation}
        k=\alpha u^2 \qquad\mbox{on~} y=0.
        \label{BoBCk0}
\end{equation}
\item
In addition, the turbulent length, $\ell\propto k^{3/2}/\epsilon$, should
be of the same order of magnitude as the boundary roughness no matter what
the mean velocity.  Hence we seek to apply
\begin{equation}
        \epsilon=\beta u^3 \qquad\mbox{on $y=0$}
        \label{BoBCe0}
\end{equation}
for some $\beta\propto 1/\ell_0$ where $\ell_0$ is the boundary roughness
length.
\item
The horizontal velocity at bottom is related to the stress via
        \begin{equation}
                \pard uy=a_u u^2\qquad\mbox{on $y=0$},
                \label{ubot}
        \end{equation}
where $a_u$ is some constant times the sign of $u$.
\end{itemize}

\section{Construction of a low-dimensional model}

In this paper we consider flows that vary ``slowly'' in $x$ and $t$.
In particular, the derivatives of the flow variables in the $x$
and $t$ directions are small quantities that can be collected with the
nonlinear part of the equations and treated as perturbations. Hence we
rewrite the equations as
\bean
\left[\matrix{0\cr
\dot u\cr \dot v\cr \dot\eta\cr \dot k\cr \dot\epsilon}\right]&=&
\left[\matrix{0 &0 &\pard{~}{y} & 0 & 0 &0\cr 0
&\pard{~}{y}\left(\nu\pard{~}{y}\right) & 0 & 0 & 0& 0\cr -\pard{~}{y} &
0&2\pard{~}{y}\left(\nu\pard{~}{y}\right) & 0 &-\frac23\pard{~}{y}& 0\cr 0 &
0&0&0&0&0\cr 0 &
0&0&0&{1\over\sigma_k}\pard{~}{y}\left(\nu\pard{~}{y}\right) &0\cr
0 & 0&0&0&0&
{1\over\sigma_\epsilon}\pard{~}{y}\left(\nu\pard{~}{y}\right)}\right]
\left[\matrix{p\cr u\cr v\cr \eta\cr k\cr \epsilon}\right]
\\ &&
+\left[\begin{array}{c} \pard{u}{x}
\\
-u\pard{u}{x}-v\pard{u}{y}-\pard{p}{x}+g\sin\theta-\frac23\pard{k}{x}
+2\pard{}{x}\left(\nu\pard{u}{x}\right)+\pard{}{y}
\left(\nu\pard{v}{x}\right)
\\
-u\pard{v}{x}-v\pard{v}{y}-g\cos\theta+\pard{}{x}\left[\nu\left(
\pard{u}{y}+\pard{v}{x}\right)\right]
\\
-u(x,\eta,t)\pard{\eta}{x}+v(x,\eta,t)
\\
-u\pard{k}{x}-v\pard{k}{y}+\left[
\pard{}{x}\left({\nu\over\sigma_k}\pard{k}{x}\right)\right]
+P_h-\lambda\epsilon
\\
-u\pard{\epsilon}{x}-v\pard{\epsilon}{y}+ \left[
\pard{}{x}\left({\nu\over\sigma_\epsilon}\pard{\epsilon}{x}\right)\right]
+ C_{\epsilon1}{\epsilon\over k}P_h
-C_{\epsilon2}\lambda{\epsilon^2\over k}
\end{array}\right]
\\
&=& \cL_{\nu}(p, \u) +\cF(p,\u,\lambda).
\eean
Treating the horizontal-space variations and the nonlinear parts
as small, one sees that $\cL_{\nu}(p,\u)$ comprises the leading term in the
equation.

With a little adaptation, the operator $\cL_{\nu}$ has critical
modes, identified by zero eigenvalues, corresponding to the variables $u$,
$\eta$, $k$ and $\epsilon$. To ensure that the turbulent energy $k$ and the
turbulent dissipation $\epsilon$ are critical modes and thus retained in
our model of turbulent floods, we introduce artificial parameters
$\lambda$ (shown in the above equation) and $\gamma$.  Making $\lambda$
small we initially neglect
the natural turbulent decay, but when $\lambda=1$ we recover the standard
$k$-$\epsilon$ model. The boundary conditions Eq.~\ref{BoBCk0} and
Eq.~\ref{BoBCe0} are, in conjunction with vertical turbulent mixing,
inherently dissipative. Thus we modify them to the following
\begin{eqnarray}
        (1-\gamma )\pard{k}{y}=\frac{\gamma}{\eta} \left(k
        - \alpha u^2\right) \quad\mbox{and~}
        \quad
        (1-\gamma )\pard{\epsilon}{y}=\frac{\gamma}{\eta} \left(\epsilon
        - \beta u^3\right) \qquad\mbox{on~} y=0,
        \label{BoBC}
\end{eqnarray}
where $\gamma$ is small in the asymptotic scheme, but eventually will be
set to $1$ to recover the desired boundary
conditions Eq.~\ref{BoBCk0} and Eq.~\ref{BoBCe0}.

\subsection{The vertical mixing operator}

The operator $\cL_{\nu}$ can be considered as determining the dominant
features of the evolution of the $k$-$\epsilon$ model Eq.~\ref{EAKeFW}. It in
turn is primarily composed of the differential operator
$$
     \pard{~}{y}\left(\nu\pard{~}{y}\right)\,.
$$
This differential operator represents vertical diffusion by turbulent
eddies.  By identifying this as the dominant term in the equations we are
physically supposing that the turbulent mixing is stronger than the
other processes that redistribute momentum and turbulent energy.
Under the conservative boundary condition
$$
   \pard{u}{y}=0\quad\mbox{for $y=0$ and $y=h$,}
$$
zero is an eigenvalue of $\pard{~}{y}\left(\nu\pard{v}{y}\right)$ with an
eigenfunction which is constant in the vertical, but which is any function
of the horizontal variable $x$. From the special structure of $\cL_{\nu}$
we deduce that it has zero as an eigenvalue, of multiplicity $4$,
corresponding to the eigenspace
$$
        (p,\u)=(g(h-y),U(x,t),0,H(x,t),K(x,t),E(x,t)),
$$
where $U$, $H$, $K$ and $E$ are arbitrary functions of $x$ and $t$. Note
that the turbulent diffusion coefficient, $\nu$, then also varies slowly in
$x$ and $t$. To leading order it is
$
   \nu_0:=C_\mu{K^2/E}
$
which is independent of $y$. Consequently, the dominant operator $\cL_{\nu_0}$
becomes effectively linear with all eigenvalues except zero being negative.
Since all other modes decay exponentially quickly, the long time behaviour
of the flow is determined by the functions $U(x,t)$, $H(x,t)$, $K(x,t)$ and
$E(x,t)$ which represent the vertically averaged horizontal velocity, the
surface elevation, the vertically averaged turbulent energy, and the
vertically averaged turbulent dissipation, respectively.

\subsection{Centre manifold expansion}
\label{ss32}

Centre manifold theory and its algebraic techniques allow us to put such a
``vertical averaged'' model on a firm footing (as has recently been done
for thin fluid films\cite{Roberts94c}). Based on a linear operator
that has decaying modes, identified by negative eigenvalues, and some
critical modes, identified by zero eigenvalues, theory\cite{Car81}
asserts that there exists a low-dimensional manifold which all
(sufficiently small) solutions approach. Once on the so-called
centre-manifold, the solutions evolve slowly according to a low-dimensional
system of evolution equations---these evolution equations form the
simplified model of the original dynamics.

In this problem, the centre-manifold will be parameterized by the four
``amplitudes,'' $U$, $H$, $K$ and $E$, which are functions of $x$ and
evolve in time $t$.
Introducing a parameter $\delta^2$ to be a measure, over all relevant
space-time, of the amplitude of the solutions, then in terms of the field
variables we define the four amplitudes such that
\begin{equation}
        \avu = \delta^2 U, \quad
        \eta = h+\delta^2 H, \quad
        \avk = \delta^2 K, \quad
        \ave = \delta^3 E,\quad
        \mbox{and also}\quad
        \pard{~}{x}=\cO(\delta^2)\,.
        \label{ampdef}
\end{equation}
Throughout, an overbar denotes a vertical average over the whole fluid
depth at any $x$ and $t$. In addition, in order to ensure that the leading
order operator is just $\cL_\nu$, we scale the parameters according to
\begin{eqnarray}
        \lambda = \lambda_0\delta\,,
        \quad
        \gamma = \gamma_0\delta^2\,,
        \quad
        \theta = \theta_0\delta^4\,,
        \label{paro}
\end{eqnarray}
so that to leading order the turbulent dynamics are conservative.

Denoting the collective amplitudes by ${\bs(x,t)}=(U,H,K,E)$, we pose
the low-dimensional assumption that the evolution of the physical
variables may be expressed in terms of the evolution of the four amplitudes:
\begin{equation}
    (p,\u)=\cV(y,\bs)
    \quad\mbox{such that}\quad
    \pard{\bs}{t}=\cG(\bs)\,. \label{CenSys}
\end{equation}
In general, we cannot find these functions $\cV$ and $\cG$ exactly as
this would be tantamount to solving exactly the original equations.
Instead we seek asymptotic approximations:
\bean
     p&\sim&p_0(y,\bs)+p_1(y,\bs)\delta+\cdots;\\
     u&\sim&u_0(y,\bs)\delta^2+u_1(y,\bs)\delta^3+\cdots;\\
     v&\sim&v_0(y,\bs)\delta^3+v_1(y,\bs)\delta^4+\cdots;\\
     \eta&=&h+\eta_0(\bs)\delta^2;\\
     k&\sim&k_0(y,\bs)\delta^2+k_1(y,\bs)\delta^3+\cdots;\\
     \epsilon&\sim&\epsilon_0(y,\bs)\delta^3+\epsilon_1(y,\bs)\delta^4+\cdots;
\eean
where the amplitudes $\bs(x,t)$ evolve in space and time according to
\bean
     \pard{U}{t}&\sim&\cU_0(U,H,K,E)\delta^2+\cU_1(U,H,K,E)\delta^3+\cdots;\\
     \pard{H}{t}&\sim&\cH_0(U,H,K,E)\delta^2+\cH_1(U,H,K,E)\delta^3+\cdots;\\
     \pard{K}{t}&\sim&\cK_0(U,H,K,E)\delta^2+\cK_1(U,H,K,E)\delta^3+\cdots;\\
     \pard{E}{t}&\sim&\cE_0(U,H,K,E)\delta^2+\cE_1(U,H,K,E)\delta^3+\cdots.
\eean
Substituting these into Eq.~\ref{EAKeFW}\footnote{For better book-keeping we
multiply the vertical momentum equation by $\delta$ and divide the energy
dissipation equation by $\delta$.} and the boundary conditions, then
collecting the coefficients of $\delta^n$, $n=0,1,\ldots$, we obtain
equations for the terms in the expansions.

As described in a number of applications\cite{Rob88,MR90,Rob2}, the
resultant hierarchy of equations may be solved. From the
equations at order $\delta^3$ we deduce
\[
    u_0=U(x,t),\quad v_0=0,\quad\eta_0=H(x,t),\quad
    k_0=K(x,t),\quad \epsilon_0=E(x,t),
\]
which introduces the critical modes of $\cL_{\nu_0}$ into the asymptotic
expansion, where $\nu_0=C_\mu K^2/E$. These predict, for example, that the
horizontal velocity is constant in the vertical. In experiments and the
environment the velocity profile is generally accepted to be approximately
logarithmic (see Fig.~12 by Shiono \& Knight\cite{Shiono91} for example).
The constant
profile here is just a leading approximation; higher orders will take
account of the drag of the bed through Eq.~\ref{ubot} to predict a curving
profile that should be closer to that observed. Our analysis here is
more systematic than others who depth-average the Navier-Stokes equations
upon assuming a constant profile as reported by Prokopiou {\em et
al}\cite[p670]{Prokopiou91}. Here all the higher-order effects combine
to jointly determine refinements to both the above constant vertical
structure and the evolution of the amplitudes.

At order $\delta^4$ the solvability condition for the inversion of
$\cL_{\nu_0}$ supplies approximate evolution equations for the
amplitudes of the low-dimensional model, namely:
\begin{eqnarray}
        \frac{\partial U}{\partial t}
        &=&\delta^2 \left(- g H_{x}+g \theta_0\right)\,,
        \label{mdl0u} \\
        \frac{\partial H}{\partial t}&=&-\delta^{2} hU_{x}\,,
        \label{mdl0h} \\
        \frac{\partial K}{\partial t}&=&-\delta^{2} E \lambda_0 \,,
        \label{mdl0k} \\
        \frac{\partial E}{\partial t}&=&-\delta^{2} {C_{\epsilon2}} E^{2}
K^{-1}
        \lambda_0\,,
        \label{mdl0e}
\end{eqnarray}
where the subscript $x$ denotes a partial derivative with respect to $x$.
These form a scaled version of the leading order evolution equations for
the four
amplitudes of the model.  Multiplying Eq.~\ref{mdl0h} by $\delta^2$ we
rewrite it as

\begin{equation}
        \frac{\partial\eta}{\partial t}+h\frac{\partial \avu}{\partial x}=0
        \label{mdl1h}
\end{equation}
which is a linear description of the conservation of water. Similarly,
multiplying Eq.~\ref{mdl0u} by $\delta^2$ we rewrite it as
\begin{equation}
        \frac{\partial\avu}{\partial t}=-g\frac{\partial \eta}{\partial x}
        +g\theta
        \label{mdl1u}
\end{equation}
which describes horizontal acceleration of water due to the slope of the
fluid surface and of the ground.
Lastly, multiplying Eq.~\ref{mdl0k} by $\delta^2$
and Eq.~\ref{mdl0e} by $\delta^3$, then setting the artificial parameter
$\lambda=1$, we rewrite them as
\begin{equation}
        \frac{\partial\avk}{\partial t}=-\ave
        \quad\mbox{and}\quad
        \frac{\partial\ave}{\partial t}=-{C_{\epsilon2}}\frac{\ave^2}{\avk}
        \label{mdl1ke}
\end{equation}
which describe a long-term algebraic decay of turbulent energy and
turbulent dissipation. Eqs.~\ref{mdl1h}--\ref{mdl1ke} form a basic,
unrefined model of the dynamics. Involving just depth-averaged
quantities they are of much lower-dimension and are much easier to solve
than the original turbulent, free-surface flow Eq.~\ref{EAKeFW}.

However, the above leading-order model Eqs.~\ref{mdl1h}--\ref{mdl1ke} omits
a number of physically important effects; for example, it does not contain
terms expressing the generation of turbulent energy through velocity shear
nor interaction with the ground. To obtain such terms we need to compute
the asymptotic expansions to higher order. Arbitrarily high order terms may
be computed in the same manner; indeed we have written a {\sc reduce}
program to compute and check the asymptotic expansions.\footnote{In some
applications\cite{MR90,Rob2,Watt} such routine computations can be
performed to 30th order and are used to show the convergence or otherwise
of the series expansions.}
For $\lambda=\gamma=1$ in the boundary conditions, the evolution on the
manifold is described up to seventh order by the equations
\bea\label{dUdt}
 \pard{\avu}{t}&=&
       \left(g\theta-g\eta_x\right)\left(1-\sfrac4{45}a_u^2\eta^2\avu^2\right)
        - {g\over3}\eta^2\eta_{xxx}
        - \sfrac23a_u\tilde\nu{\avu^2\over \eta}\left(1-a_u\eta\avu\right)
      \\\nonumber&&
       - \avu\pard{\avu}{x}
       + 2\pard{~}{x}\left(\tilde\nu\pard{\avu}{x}\right)\,,
\\\label{dHdt}
 \pard{\eta}{t}&=&-\pard{~}{x}(\avu\eta)\,,
\\\label{dKdt}
 \pard{\avk}{t}&=&-\ave-{4\tilde\nu\over3\sigma_k\eta^2}\avk
           +\frac{\tilde\nu\alpha}{\sigma_k\eta^2}{\avu^2}
           -\avu\pard{\avk}{x}
           +{1\over
{\sigma_k}}\pard{~}{x}\left(\tilde\nu\pard{\avk}{x}\right)\,,

\\\label{dEdt}
 \pard{\ave}{t}&=&-{C_{\epsilon2}} {\ave^{2}\over \avk}
                -{4\tilde\nu\over3\sigma_\epsilon \eta^2}\ave
                +{\tilde\nu\beta\over\sigma_\epsilon \eta^2}\avu^3
                - \avu\pard{\ave}{x}
                + {1\over \sigma_\epsilon}\pard{~}{x}\left(
                    \tilde\nu\pard{\ave}{x}\right)\,,
\eea
where the model turbulent diffusivity is $\tilde\nu=C_\mu\avk^2/\ave$.

The terms in these equations are all physically reasonable. The $\avk$ and
$\ave$ equations have the same form and respectively the terms represent:
basic interaction between turbulent energy and dissipation; decay due to
turbulent transport to and consequent loss across the bed; production due
to velocity shear in the vertical; advection at the average velocity; and
horizontal turbulent dispersion. The $\avu$ equation contains terms
representing, respectively: accelerations due to surface and bed slope,
somewhat ameliorated by a small correction in $a_u\eta\avu$; wave
dispersion; dissipation which is quadratic in the velocity,\footnote{Recall
that $a_u$ is a constant times the sign of $\avu$, so that this quadratic
term is always a dissipative drag.} also ameliorated by a correction in
$a_u\eta\avu$; advection; and horizontal turbulent dispersion.
Interestingly, the turbulent energy density and dissipation only enter
directly the $\avu$ equation through the eddy viscosity $\tilde\nu$.

We expect that this set of equations has wide applicability in the
modelling of turbulent floods and other shallow water flows where
turbulence is significant.

\section{Recovery of  classical models}

In the absence of any slope, the leading order
model Eqs.~\ref{mdl1h}--\ref{mdl1u} reduces to the linearised long-wave
equation
$$
\frac{\partial^2\eta}{\partial t^2}=gh\frac{\partial^2\eta}{\partial
x^2}\ .
$$
Similarly, selected versions of the higher-order modifications reduce
to other traditional equations modelling the long-wave dynamics of water as
we now record.

Examining the model dynamical system, one sees the {\em St Venant
equations\/} for the flow in an open rectangular channel follow from the
low orders of Eqs.~\ref{dUdt}--\ref{dHdt} when the turbulent energy $\bar
k$ and turbulent dissipation $\bar\epsilon$ are assumed constant in $x$
and $t$:
\bea
   \pard{\eta}{t}+\avu\pard{\eta}{x}+\eta\pard{\avu}{x}&=&0,
\label{SV1}\\
   \pard{\avu}{t}+\avu\pard{\avu}{x}+g\pard{\eta}{x}&=&g
   \theta-C_f{\avu^2\over \eta}\,.
   \label{SV2}
\eea
We deduce the friction coefficient to be $C_f=\frac23a_u\tilde\nu$ in terms of
the parameters of the $k$-$\epsilon$ model and the underlying turbulent
diffusivity $\tilde\nu$.

Similarly, the {\em Boussinesq equations\/}
\bea
    \pard{\eta}{t}+\pard{(\avu{\eta})}{x}&=&0,          \label{BS1}\\
   \pard{\avu}{t}+\avu\pard{\avu}{x}+ g\pard{\eta}{x}+\frac13gh^2
   {\partial^3\eta\over\partial x^3}&=&0,       \label{BS2}
\eea
are recovered when: the level of turbulence is low (small $\tilde\nu$)
either through minimal production of turbulence in Eqs.~\ref{dKdt}--\ref{dEdt}
or through little bed friction (small $a_u$); the ground does not slope
($\theta=0$); and, except for dispersion ($\eta_{xxx}$), high-order effects
are neglected.

\paragraph{Acknowledgement}
This research was supported by a grant from the Faculty of Sciences of
the University of Southern Queensland, and by the Australian Research
Council.

\end{document}